\documentclass[twocolumn,prb,showpacs,multicol,amsmath,amssymb]{revtex4}
\usepackage[dvips]{graphicx}
\usepackage{graphics}
\usepackage{color}
\def\be{\begin{equation}}
\def\ee{\end{equation}}

\def\a{\alpha}
\def\b{\beta}

\def\s{\sigma}
\def\d{\dagger}
\def\bi{\begin{itemize}}
\def\ei{\end{itemize}}
\def\bn{\begin{enumerate}}
\def\en{\end{enumerate}}
\def\bea{\begin{eqnarray}}
\def\eea{\end{eqnarray}}
\def\no{\nonumber}
\def\ba{\begin{array}}
\def\ea{\end{array}}
\def\bd{\begin{displaymath}}
\def\ed{\end{displaymath}}
\def\la{\langle}
\def\ra{\rangle}
\begin{document}

\title{A signature for the Luttinger liquid phase in alternating Heisenberg spin-1/2 chains}

\author {J. Abouie$^{1,2}$ and S. Mahdavifar$^{3}$ }

\affiliation{
$^{1}$Physics Department, Shahrood University of Technology, Shahrood 36199-95161,
Iran\\
 $^{2}$School of physics, Institute for Research in Fundamental Sciences, IPM 19395-5531 Tehran, Iran\\
$^{3}$Department of Physics, University of Guilan, P.O.Box 41335-1914,
Rasht, Iran}

\begin{abstract}
We have studied the zero and low-temperature
behavior of anisotropic alternating
antiferromagnetic-ferromagnetic Heisenberg spin-1/2 chains in a
transverse magnetic field. Using the analytical spinless fermion
approach, the thermodynamic behavior of the model has been
studied.  We have introduced a new order parameter to distinguish
the gapless Luttinger liquid phase from the other gapped phases. The exact diagonalization Lanczos results are also used to compare with the spinless fermion ones.
We have found a double peak structure in the specific heat curves
for the region between the two quantum critical points. Using the numerical full diagonalization results, the existence of the
double peak structure is confirmed.
 \end{abstract}
\maketitle

\section{introduction}

The study of continuous phase transitions has been one of the most
fertile branches of theoretical physics in the last decades. Each
phase can usually be characterized by an order parameter. Often,
the choice of an order parameter is obvious. However, in some
cases finding an appropriate order parameter is complicated. In
particular, many magnetic systems experience the Luttinger liquid
(LL) phase for certain values of some non-thermal control
parameter. Luttinger liquid is the paradigm for the description of
interacting one-dimensional (1D) quantum systems\cite{Giam04,
Taka99}. The correlation functions decay as power laws and the
ground state has a quasi-long range order. In spite of all
manifestations of this phase, the magnetic ordering in the LL
phase is still unclear, i.e, all suggested local magnetic order
parameters turn out to be zero, having no effect in determining
the structure of the LL phase. In this work, we will concentrate
on studying the probably nonlocal magnetic ordering of the LL
phase.

One dimensional quantum spin systems such as antiferromagnetic
spin-1/2 chains\cite{Lake05}, spin ladder systems\cite{Dago96}, or
bond alternating spin-1/2 AF-F chains are good candidates
for studying the LL phase. When these systems are placed in an
external magnetic field, they can be mapped onto a 1D system of
interacting spinless fermions\cite{Giam04, Chit97, Usam98,Furu99,
Hiki01, Yama05}. The filling of a fermionic band can thus be
continuously tuned, making these systems suitable for probing the
LL physics\cite{Giam04, Lake05, Okun07, Kimu08}. In the case of
isotropic bond alternating AF-F spin-1/2 chains the whole band can
be covered. In the absence of an external magnetic field ($h=0$),
the model is mapped onto a nonlinear sigma model with a $4\pi s$
topological angle\cite{Bosq00}. This model is always gapful and
can be regarded as a Haldane gapped spin-1 chain\cite{Hald83}. The
gap decreases by increasing the magnetic field and goes to zero at
the lower critical field $h_{c_{1}}$\cite{Mahd1-08}. At $h_{c_1}$,
the system enters the gapless phase and the fermionic band starts
to be filled. This process continues until the upper critical
field $h_{c_2}$ is reached. Increasing the field beyond $h_{c_2}$
reopens the gap and the band is then completely filled.

Recently, It was demonstrated that ${\rm CuBr_4(C_5H_{12}N)_2}$ is
a unique system for controlling and probing the physics of LL
\cite{Klan08}. This sample is a spin ladder system with
$h_{c_1}=6.6$ T and $h_{c_2}=14.6$ T\cite{Wats01}. Bond
alternating AF-F spin-1/2 chains that allow experimental access to
the whole fermionic band are not known. However the
$\rm{(CH_3)_2NH_2CuCl_3}$ system has been considered to be a
suitable realization of bond alternating AF-F spin chain.
Linked-cluster calculations and bulk measurements show that
$\rm{DMACuCl_3}$ is also a realization of the spin-1/2 alternating
AF-F chain with nearly the same strength of antiferromagnetic and
ferromagnetic couplings\cite{Ston07}. Other experimental samples
of the AF-F alternating spin-1/2 chain compounds have also been
reported in
Refs.[\onlinecite{Hagi97,Taka97,Koda99,Mana97-1,Mana97-2}].
Although it is not known whether these systems can experience the
LL phase, theoretically, an isotropic bond alternating AF-F spin
chain, in its ground state, can enter the LL phase with quasi-long
range order. In this paper, we look for an order parameter to
describe the LL phase of the bond alternating AF-F spin-1/2 chain.
The Hamiltonian of this model with anisotropic ferromagnetic
coupling is given by
\begin{eqnarray}
\no\hat{H}&=&-J_F\sum_{j=1}^{N/2}(S_{2j}^xS_{2j+1}^x+
\Delta S_{2j}^yS_{2j+1}^y+ S_{2j}^zS_{2j+1}^z)\\
&&+J_{AF}\sum_{j=1}^{N/2}{\bf S}_{2j-1}\cdot{\bf
S}_{2j}+h\sum_{j=1}^N S_j^{\alpha}, \label{H}
\end{eqnarray}
where $S_{j}^{\alpha}, (\alpha=x, y, z)$ are spin-1/2 operators on
the $j$-th site. $J_{F}$ and $J_{AF}$ denote the ferromagnetic and
antiferromagnetic couplings, respectively. $\Delta$ is the
anisotropy parameter and $h$ is a uniform magnetic field.

The ground-state
properties\cite{Taka92,Hida92a,Hida93,Saka95,Hida92b,
Kohm92,Yama93} and low-lying excitations\cite{Hida94} of this
model have been thoroughly investigated by numerical tools and
variational schemes. In particular, the string order parameter
which was originally defined for the spin-1 Heisenberg
chains\cite{Nijs89} has been generalized to this system. The
ground state has long-range string order, which is characteristic
of the Haldane phase. Hida has shown that the Haldane phase of the
AF-F alternating chain is stable against any strength of
randomness\cite{Hida99}. The ground state phase diagram of the
AF-F alternating chain in a longitudinal ($\alpha=y$) magnetic
field is studied using numerical diagonalization and finite-size
scaling based on conformal field theory\cite{Saka95}. It is shown
that the magnetic state is gapless and described by the LL phase.
The model represented by the Hamiltonian in Eq. \ref{H} which
includes a transverse magnetic field has been studied recently
using the numerical Lanczos method\cite{Mahd1-08}. The main
attention in this study was focused on the investigation of
field-induced effects in the ground state phase diagram, present
when the antiferromagnetic coupling $J_{AF}$ dominates
($J_{AF}>J_{F}$). The system has two critical fields and the
energy gap in the intermediate region depends on the anisotropic
parameter $\Delta$. For $\Delta\neq1$, the intermediate state is
gapful and the ground state of the model has
stripe-antiferromagnetic order\cite{Mahd1-08}.

In this paper we consider again an anisotropic AF-F chain in a
transverse magnetic field. Using the numerical exact
diagonalization method and the analytical spinless fermion
approach, we investigate the zero temperature and thermodynamic
behavior of the model. We introduce a new mean field order
parameter which can distinguish the LL phase from the other gapped
phases (Fig. \ref{fig2}). In the specific heat curves versus
temperature, a double peak structure appears in the intermediate
region of the magnetic fields $h_{c_1}<h<h_{c_2}$.

The outline of the paper is as follows. In section II we discuss
the zero-temperature ground state phase diagram of the model. In
section III we present the results of the spinless fermion
approach and the numerical full diagonalization results on the
low-temperature behavior of the model. Finally we conclude and
summarize our results in section IV.


\section{zero-temperature behavior}
\subsection{RESULTS FROM THE NUMERICAL LANCZOS METHOD}
In this section we briefly discuss the model (1) in the limiting
case of the strong AF coupling $J_{AF}\gg J_{F}$. In this limit
the model can be mapped onto an effective spin-chain
Hamiltonian\cite{Mila98}. At $J_{AF}\gg J_{F}$, the system behaves
as a nearly independent block of pairs\cite{Mila98}. Indeed an
individual block may be in a singlet
or a triplet state
with the corresponding energies given by
$$
E_{1, -1}={J_{AF}\over 4}\ \mp h,\,\, E_0={J_{AF}\over 4},\,\,
E_s=-{3J_{AF}\over 4}.
$$
For $h\leq J_{AF}$, one component of the triplet becomes closer to
the singlet ground state such that for a strong enough magnetic
field we have a situation when the singlet and $S^{z} =1$
component of the triplet create a new effective spin $\tau=1/2$
system. On the new singlet-triplet subspace, the original
Hamiltonian  becomes the Hamiltonian of a fully anisotropic XYZ
spin-1/2 chain in an effective magnetic field\cite{Mahd1-08}
\begin{eqnarray}
H_{\emph{eff}} &=& \frac{J_{F}}{2} \sum_{j=1}^{N/2}
[-\frac{1}{2}\tau^{z}_{j}\tau^{z}_{j+1} + \Delta \tau^{y}_{j}
\tau^{y}_{j+1} +\tau^{x}_{j} \tau^{x}_{j+1}] \nonumber \\
&+& h^{eff} \sum_{j=1}^{N/2} \tau^{z}_{j}\, ,
\label{EffectiveHamiltonian2}
\end{eqnarray}
where $\mathrm{h}^{eff} = h-J_{AF}+J_{F}/4$. At $\Delta = 1$, the
effective problem reduces to the theory of the $XXZ$ chain with a
fixed antiferromagnetic anisotropy of $1/2$ in a magnetic
field\cite{Taka99}. The gapped Haldane phase at
$\mathrm{h}^{eff}<\mathrm{h}^{eff}_{c_1}=-\frac{J_F}{4}$ for the
AF-F alternating chain corresponds to the negatively saturated
magnetization phase for the effective spin chain, whereas the
massless LL phase of the AF-F alternating chain corresponds to the
finite magnetization phase of the effective spin-1/2 chain. The
critical field $\mathrm{h}^{eff}_{c_2}=\frac{J_F}{4}$ where the
AF-F alternating chain is totally magnetized, corresponds to the
fully magnetized phase of the effective spin chain.

Away from the isotropic point $\Delta =1$ the effective
Hamiltonian (\ref{EffectiveHamiltonian2}) describes the fully
anisotropic ferromagnetic XYZ chain in a magnetic field that is
directed perpendicular to the easy axis. In this case, it is
found\cite{Mahd1-08} that a gapped stripe-antiferromagnetic phase
exists for the intermediate values of the transverse magnetic
field $h_{c_1}<h<h_{c_2}$.

\begin{figure}
\centerline{\includegraphics[width=9cm,angle=0]{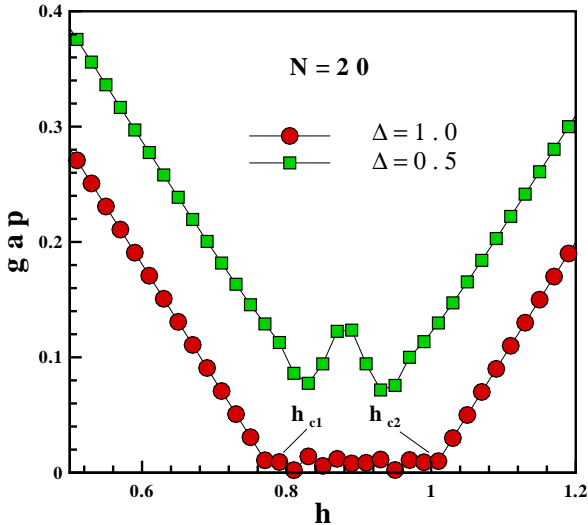}}
\caption{The excitation gap of a spin-1/2 AF-F chain versus the
uniform  magnetic field $h$, for different anisotropy parameters
$\Delta=1.0, ~0.5$  and chain size $N=20$.} \label{fig1}
\end{figure}

To recognize the different phases induced by the transverse
magnetic field in the ground-state phase diagram, we implemented
the modified Lanczos algorithm of finite-size chains $(N=12, 16,
20, 24)$\cite{Mahd1-08} with $J_{AF}=1, J_F=1/2$ and different
values of the anisotropy parameter $\Delta$. The energies of a few
lowest eigenstates were obtained for the chains with periodic
boundary conditions. In Fig.~\ref{fig1} we have plotted the
results of these calculations for different values of the
anisotropy parameter $\Delta=1.0,~0.5$, and chain length $N=20$.
As it is clearly seen from this figure, in the case of zero
magnetic field the spectrum of the model is gapped. For $h\neq 0$
the gap decreases linearly with $h$ and vanishes at the critical
field $h_{c_{1}}(\Delta)$. In the isotropic case $\Delta=1.0$, the
spectrum remains gapless for $h_{c_{1}}=0.78 \pm
0.01<h<h_{c_{2}}=1.0$ and becomes once again gapped for
$h>h_{c_{2}}$. But, in the anisotropic case $\Delta=0.5$, the
excitation spectrum is gapfull except at two critical field values
$h_{c_{1}}=0.82\pm0.01$ and $h_{c_{2}}=0.94\pm0.01$. In the
intermediate region $h_{c_{1}}<h<h_{c_{2}}$ the spin gap which
appears at $h>h_{c_{1}}$, first increases vs external field and
after passing a maximum decreases to vanish at $h_{c_{2}}$.

In conclusion, at $T=0$, two quantum phase transitions in the
ground-state phase diagram of the model have been identified with
increasing transverse magnetic field \cite{Mahd1-08}. The first
transition corresponds to the transition from the gapped Haldane
phase to the gapless LL phase (or gapped stripe-antiferromagnetic
for the anisotropic case). The other one is the transition from
the gapless LL phase (or gapped stripe-antiferromagnetic) to the
fully polarized phase.

\subsection{FERMIONIZATION}

The behavior of the gap and the quantum phase space of the model
lead us to investigate the thermodynamic properties of the model.
In this respect, we implement the Jordan-Wigner transformation to
fermionize the model. Because of two types of coupling constant,
we introduce two kinds of spinless fermion through the following
Jordan-Wigner transformations:
\begin{eqnarray}
\no S_{2n-1}^+&=&a_n^{\d}e^{i\pi\bigg(\sum_{m=1}^{n-1}a_m^{\d}a_m+\sum_{m=1}^{n-1}b_m^{\d}b_m\bigg)},\\
\no S_{2n}^+&=&b_n^{\d}e^{i\pi\bigg(\sum_{m=1}^{n}a_m^{\d}a_m+\sum_{m=1}^{n-1}b_m^{\d}b_m\bigg)},\\
\no
S_{2n-1}^{z}&=&a_n^{\d}a_n-\frac{1}{2},~~~~~~~S_{2n}^{z}=b_n^{\d}b_n-\frac{1}{2}.
\label{JW}
\end{eqnarray}
Using the above transformations, the thermodynamic behavior of the
isotropic ($\Delta=1$) Heisenberg AF-F spin-1/2 chains has been
studied in the absence of a magnetic field\cite{Yama05}. By the
above transformations, the AF-F spin chain is mapped onto a 1D
system of interacting spinless fermions;
\begin{eqnarray}
\no\hat{H}_f&=&-\frac{N
h}{2}+\sum_{n=1}^{N/2}(\frac{J_F}{2}-\frac{J_{AF}}{2}+h)(a_n^{\d}a_n+b_n^{\d}b_n)\\
\no&&+\sum_{n=1}^{N/2}(\frac{J_{AF}}{2}a_n^{\d}b_n-\frac{1+\Delta}{4}J_Fa_{n+1}^{\d}b_n\\
\no&&-\frac{1-\Delta}{4}J_F b_n^{\d}a_{n+1}^{\d}+h.c)
+\sum_{n=1}^{N/2}J_{AF}a_n^{\d}a_nb_n^{\d}b_n\\
&&-\sum_{n=1}^{N/2}J_Fb_n^{\d}b_na_{n+1}^{\d}a_{n+1}.
\end{eqnarray}
Treating the Hamiltonian $H_f$ in the mean field approximation,
the interacting fermionic system reduces to a 1D system of new
non-interacting dynamical quasi-particles. Many mean field order
parameters (auxiliary fields) might be considered. Many of them
are irrelevant, \emph{i.e.}, they do not have a stable mean field
solution, while some are relevant. In our system we have
introduced the magnetization, ferromagnetic and antiferromagnetic
dimers as mean field order parameters;
\begin{eqnarray}
\label{or-pa}\la a_n^{\d}a_n\ra&=&d_a,~~~ \la b_n^{\d}b_n\ra=d_b,\\ \nonumber
\la a_{n+1}^{\d}b_n\ra&=&P_F, ~~\la b_n^{\d}a_n\ra= P_{AF}.
\end{eqnarray}
Utilizing the above order parameters, the mean field Hamiltonian is
given by;
\begin{equation}
{\cal
H}_{HF}=E_0+\sum_k{\frac{A}{2}a_k^{\d}a_k+\frac{B}{2}b_k^{\d}b_k+\gamma_ka_k^{\d}b_k+h.c}
\label{Ha-Fo}
\end{equation}
where,
\begin{eqnarray}
\no A&=&(J_{AF}-J_F)(d_b-1/2)+h ,\\
\no B&=&(J_{AF}-J_F)(d_a-1/2)+h ,\\
\no\gamma_k&=&J_{AF}(1/2-P_{AF})e^{ik/2}+J_F(P_F^*-\frac{1+\Delta}{4})e^{-ik/2}\\
\no E_0&=&\frac{N}{2}\bigg(J_{AF}(|P_{AF}|^2-d_ad_b+1/4)-h\\
\label{param}&&-J_F(|P_F|^2-d_ad_b+1/4)-\frac{J_F}{16}(1-\Delta^2)^2\bigg).
\end{eqnarray}
In the above, $\la\dots\ra$ represents thermal averaging over the
Hartree-Fock eigen-states. $d_a$ and $d_b$ are related to the
magnetization, and $P_F$ and $P_{AF}$ are ferromagnetic and
antiferromagnetic exchange order parameters, respectively. It is
clear that $P_{AF}(P_{F})$ in the Hilbert space is the AF(F)-dimer
order parameter, \emph{i.e.},
\begin{eqnarray}
\no&&\la S^-_{2n}S^+_{2n+1}\ra=\la a_{n+1}^{\d}b_n\ra=P_F,\\
\label{ord.pa}&&\la S^-_{2n-1}S^+_{2n}\ra=\la b_n^{\d}a_n\ra= P_{AF}.
\end{eqnarray}

Using the following unitary transformations
\begin{eqnarray}
\no a_k&=&u_k\a_k+v_ke^{i\theta_k}\b_k,\\
\no b_k&=&v_ke^{-i\theta_k}\a_k+u_k\b_k,
\end{eqnarray}
the diagonalized Hamiltonian is given by
\begin{eqnarray}
{\cal
H}&=&E_0+\sum_k\sum_{\s=\pm}(\epsilon_k^+\a^{\d}\a_k+\epsilon_k^-\b^{\d}\b_k),\\
\label{Hd}&&\epsilon_k^{\pm}=\xi\pm\sqrt{\eta^2+\gamma_k^2}+h,\\
\no&&\eta=\frac{B-A}{2},~~~~~~\xi=\frac{B+A}{2},
\end{eqnarray}
Equation(\ref{param}) clearly shows that
the effect of the anisotropy appears in the dispersion relations.
The dispersion relations of low-lying excitation read as,
\begin{eqnarray}
\no\epsilon_k^{\pm}\cong\pm[(J_{AF}\tilde{P}_{AF}-J_F\tilde{P}_{F})^2+J_{AF}J_F\tilde{P}_{AF}\tilde{P}_{F}
k^2]^{\frac{1}{2}}+h,
\end{eqnarray}
where $\tilde{P}_{AF}=1/2-ReP_{AF}$ and
$\tilde{P}_{F}=\frac{1+\Delta}{4}-ReP_F$. Using the above order
parameters, the thermodynamic functions such as the internal
energy and the specific heat are expressed as follows:
\begin{eqnarray}
\no U&=&E_0+\sum_k\sum_{\s=\pm}\epsilon_k^{\s}\bar{n}_k^{\s},\\
C&=&\frac{\partial U}{\partial T},
\end{eqnarray}
where, $\bar{n}_k^{\s}=[e^{\frac{\epsilon_k^{\s}}{T}}+1]^{-1}$ is
the fermion distribution function (choose $k_{\rm B}=1$).

In order to obtain the thermodynamic behavior of the system, we
need to know the whole temperature behavior of the order
parameters for different values of $h$. The order parameters
satisfy a set of self consistent equations and it should be solved
the equations (\ref{or-pa}), self-consistently.

\begin{figure}[h]
\centerline{\includegraphics[width=8cm,angle=0]{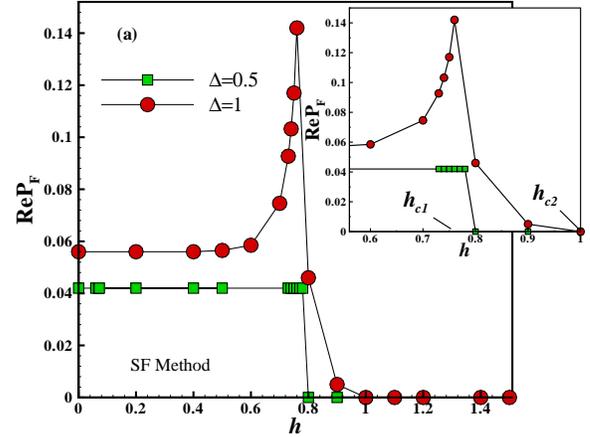}}
\centerline{\includegraphics[width=8cm,angle=0]{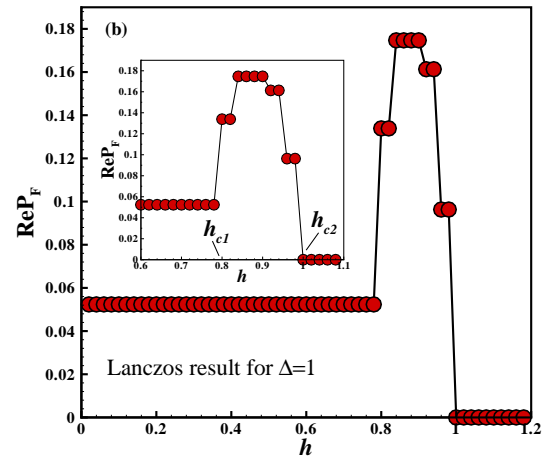}}
\caption{ a) The spinless fermion results of the real part of the
ferromagnetic exchange order parameter at temperature $T=0.05$
versus $h$, $J_{AF}=1$ and $J_{F}=0.5$, b) Exact diagonalization
Lanczos results of the real part of the $P_F$ for a chain size
$N=20$.} \label{fig2}
\end{figure}

In Fig.\ref{fig2}-(a), we have plotted the ferromagnetic dimer order
parameter of both anisotropic $(\Delta=0.5)$ and isotropic
($\Delta=1$) AF-F Heisenberg spin-1/2 chains versus $h$. The
curves have been plotted close to the zero temperature.
As we mentioned, at $T=0$ for different values of $h$, our 1D
system experiences three phases. In the Haldane phase
($h<h_{c_1}$), the quantum fluctuations are strong enough to
suppress the ferromagnetic order of the system and the AF(F)-dimer
parameters $P_{AF}(P_F)$ should be close to the classical value
$-0.5(0)$\cite{Japa07}. In this region the energy spectrum is
gapped and the gap is decreased by increasing $h$.
The first quantum phase transition occurs at $h_{c_1}$.
At the same time the fermionic band starts to be filled where the
excited state sticks to the ground state and the gap of the system
is closed. For the first region there is no specific difference
between the two anisotropic and isotropic cases. \emph{i.e.},
anisotropy shows a marked behavior for the case of
$h_{c_1}<h<h_{c_2}$.

For the case of $\Delta=1$ and $h=0$, the system is isotropic and has SU(2)
rotational symmetry. In the intermediate region this system is
gapless and correlations decay in a power law\cite{Saka95}. In
this region, the Luttinger liquid phase, the magnetic field
suppresses quantum fluctuations and induces quasi long-range order
in the 1D system. Although the power law behavior of the
correlations and other properties of the LL phase are manifest,
the lack of an order parameter to show the behavior of the system
in this phase is still felt. As it is obviously seen from
Fig.\ref{fig2}, the F-dimer order parameter has a considerable
value in this region and is sensitive to $h$. The
value of this order parameter falls sharply at the second critical
field where the fermionic band is completely filled and the gap of
the system is reopened.

We have also plotted in Fig.\ref{fig2}-(b) the F-dimer order
parameter of the isotropic AF-F Heisenberg spin-1/2 chain by
using the exact diagonalization Lanczos results. The numerical
Lanczos method is implemented as a reference approach to see the
accuracy of our spinless fermion method. As it is clearly
observed from Fig.\ref{fig2}-(b) the results of Lanczos method,
confirms that the defined mean field order parameter has
considerable value for the intermediate values of field.

We have also plotted in Fig.\ref{fig2} the F-dimer order parameter
of anisotropic F-AF spin-1/2 chain versus transverse magnetic
field. Regarding the anisotropy in ferromagnetic coupling, the
anisotropic system is mapped to a 1D XYZ model with U(1) symmetry. In the intermediate region
($h_{c_1}<h<h_{c_2}$) the system has the magnetic long range order and spins
are aligned stripe-antiferromagnetically. This phase is gapfull
and the F-dimer order has no considerable value in the
this phase. Moreover, there is a big discrepancy
between the values of the F-dimers of the isotropic and
anisotropic cases, \emph{i.e.} in the LL gapless phase the
behavior of the F-dimer order parameter ($ReP_F$) is different
from those of the gapped phases. Therefore, we can conclude that
the above F-dimer order parameter can distinguish between the
gapless LL phase and the other gapped phases.

This means that experimenters might look for such behavior in
$ReP_F$ at finite temperatures. \emph{I.e.}, experimental results
on this quantity may be used to distinguish between the isotropic
chains and anisotropic ones (gapped stripe-antiferromagnetic and
gapless LL phases).


\section{Thermal behavior of the model}

In this section we will study the thermal behavior of the mean
field order parameters of the Heisenberg spin-1/2 AF-F chains in
both anisotropic and isotropic cases. The order parameters also
have different behavior for different values of $h$.

For $h<h_{c_1}$, in the Haldane phase, an increase in temperature
increases the thermal fluctuations and Haldane ordering is
suppressed. The solution of the Eqs. \ref{or-pa} shows that the
AF-dimer order parameter ($ReP_{AF}$) decreases with temperature.

\begin{figure}[h]
\centerline{\includegraphics[width=8cm,angle=0]{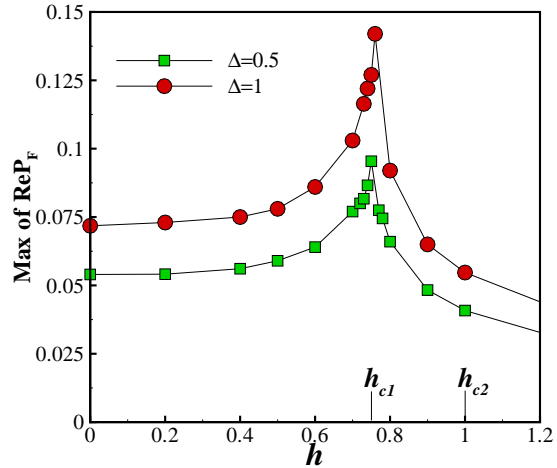}}
\caption{ The spinless fermion results of the real part of the
ferromagnetic exchange order parameter at temperature $T_F$ versus
$h$, for $J_{AF}=1$ and $J_{F}=0.5$ .} \label{fig3}
\end{figure}

However, $J_F$ couplings supply an interaction between the unit
cells and try to establish F-dimer order ($ReP_{F}$) even at
moderate temperatures (scaled with $J_F$), \emph{i.e.} ($ReP_{F}$)
increases with temperature up to $T_F$ (see Fig. \ref{fig3}).
Above $T_F$, fluctuations are large enough to decrease $ReP_F$.
The locations of the maximum value of ($ReP_F$) with respect to
$h$ are shown in Fig.\ref{fig3}. As it is obviously seen from this
plot, there is a peak at $h$ very close to the first critical
field. The location of this peak is a good candidate to find the
first critical point.

\begin{figure}[h]
\centerline{\includegraphics[width=8cm,angle=0]{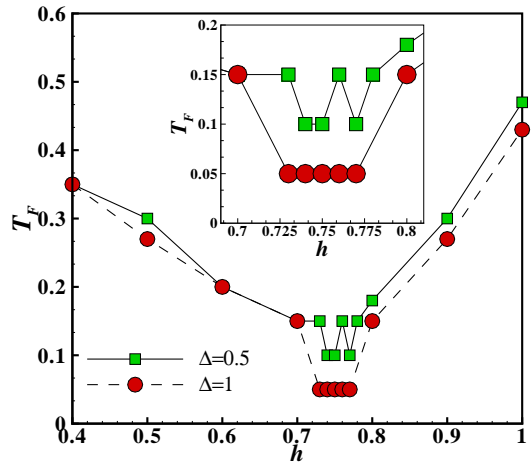}}
\caption{ The spinless fermion results for $T_F$ versus $h$, with
$J_{AF}=1$ and $J_{F}=0.5$ .} \label{fig4}
\end{figure}

By increasing the magnetic field, the value of $T_F$ (location of
the maximum value of $ReP_F$) decreases. It is found that $T_F$
goes to a minimum (Fig.\ref{fig4}) when $h$ increases up to
$h_{c_1}$. Increasing the magnetic field further and for
$h_{c_1}\sim0.76<h<h_{c_2}\sim1$, $T_F$ always maintains a minimum
value and the value of $ReP_F$ decreases monotonically with
increasing $T$. Increasing the magnetic field even further, for
$h>h_{c_2}$, $T_F$ also increases.

The SF calculations also show that the behavior of $T_F$ is
similar to the energy gap (see Fig.\ref{fig1})

\begin{figure}
\centerline{\includegraphics[width=8cm,angle=0]{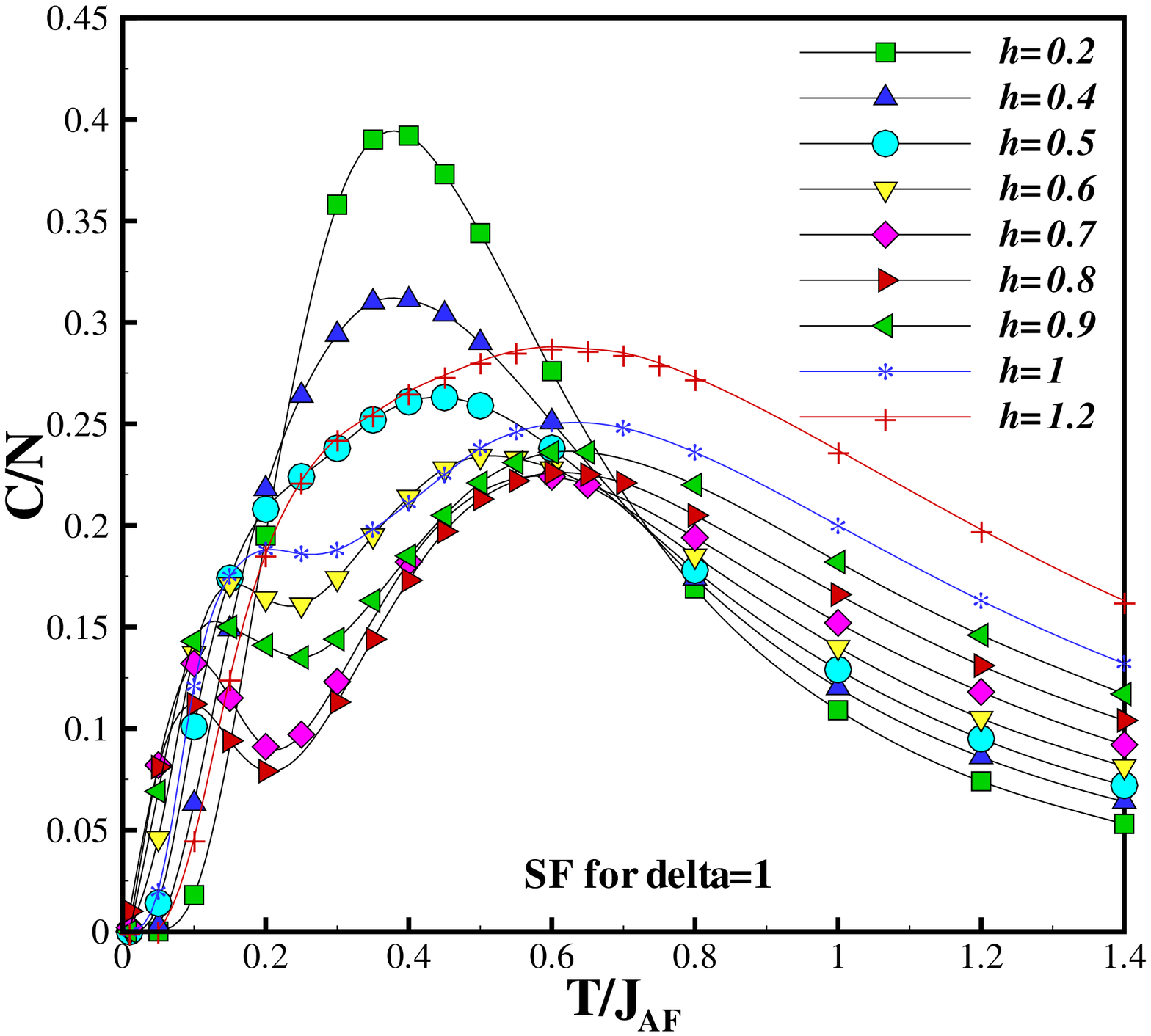}}
\centerline{\includegraphics[width=8cm,angle=0]{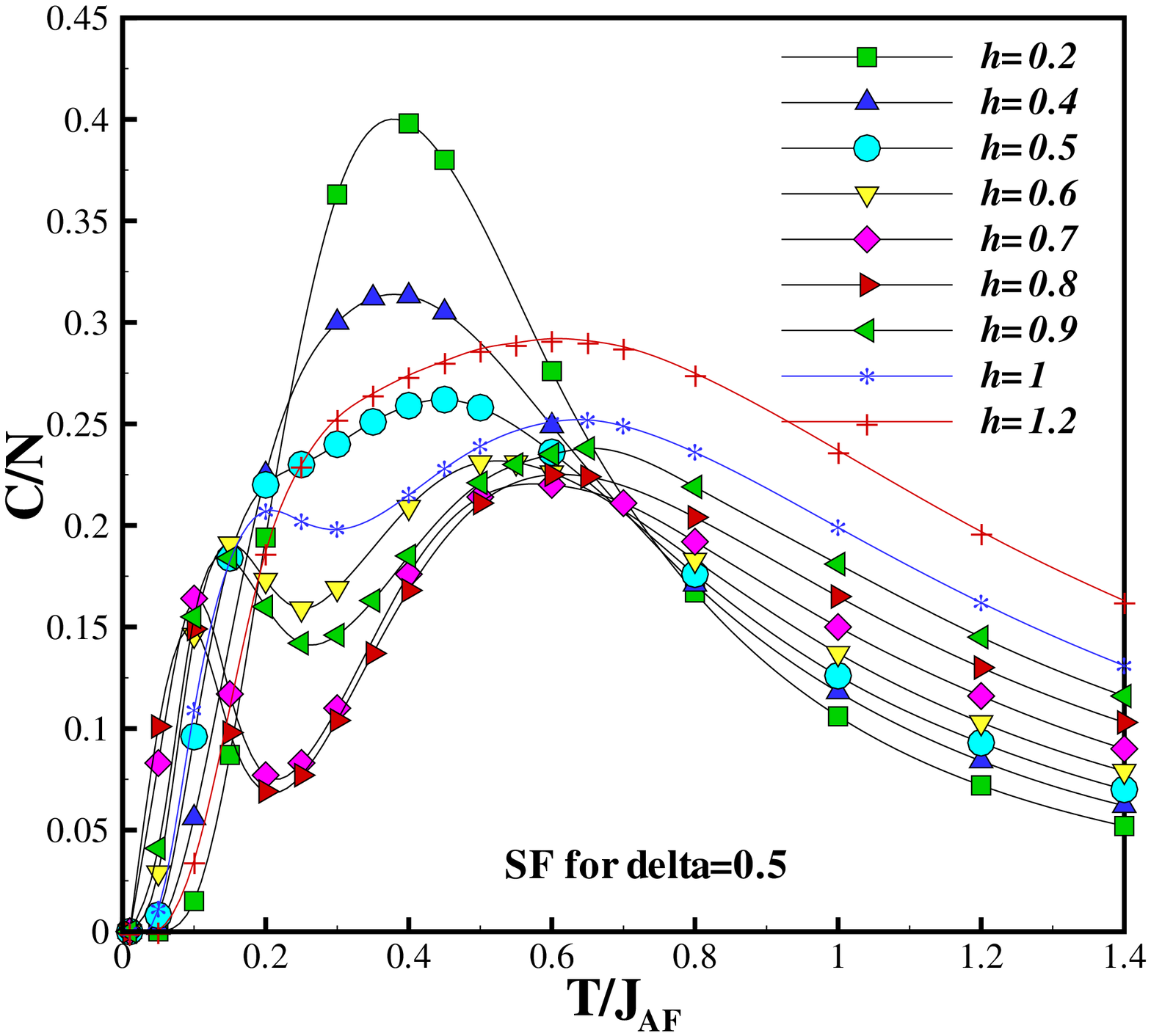}}
\caption{ The spinless fermion results of the specific heat of an
alternating AF-F spin-1/2 chain versus $T$ for different values of
the magnetic field $h$ with $J_{AF}=1$ and $J_{F}=0.5$. (a)
$\Delta=1$ (b) $\Delta=0.5$.} \label{fig5}
\end{figure}
\begin{figure}
\centerline{\includegraphics[width=8cm,angle=0]{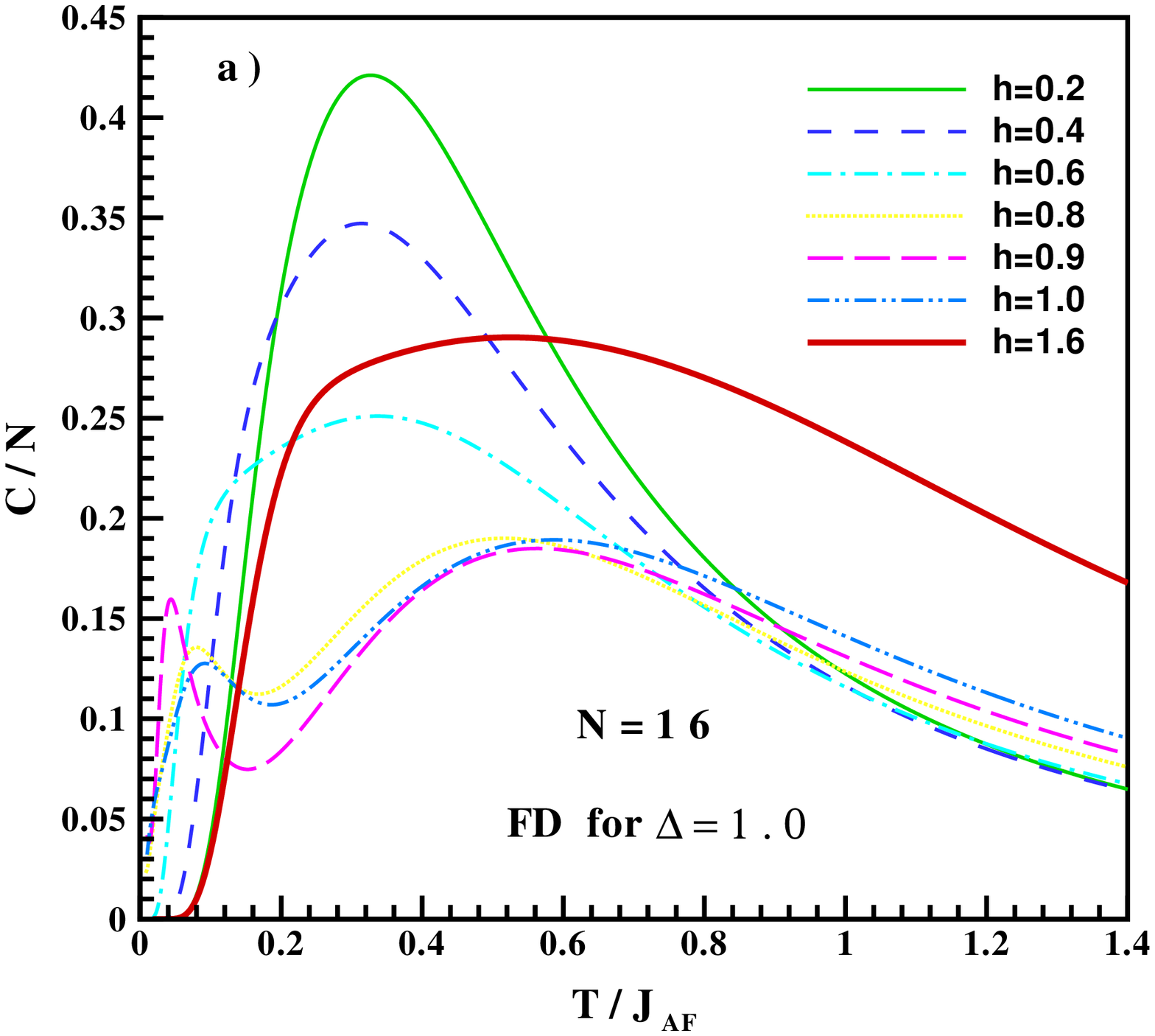}}
\centerline{\includegraphics[width=8cm,angle=0]{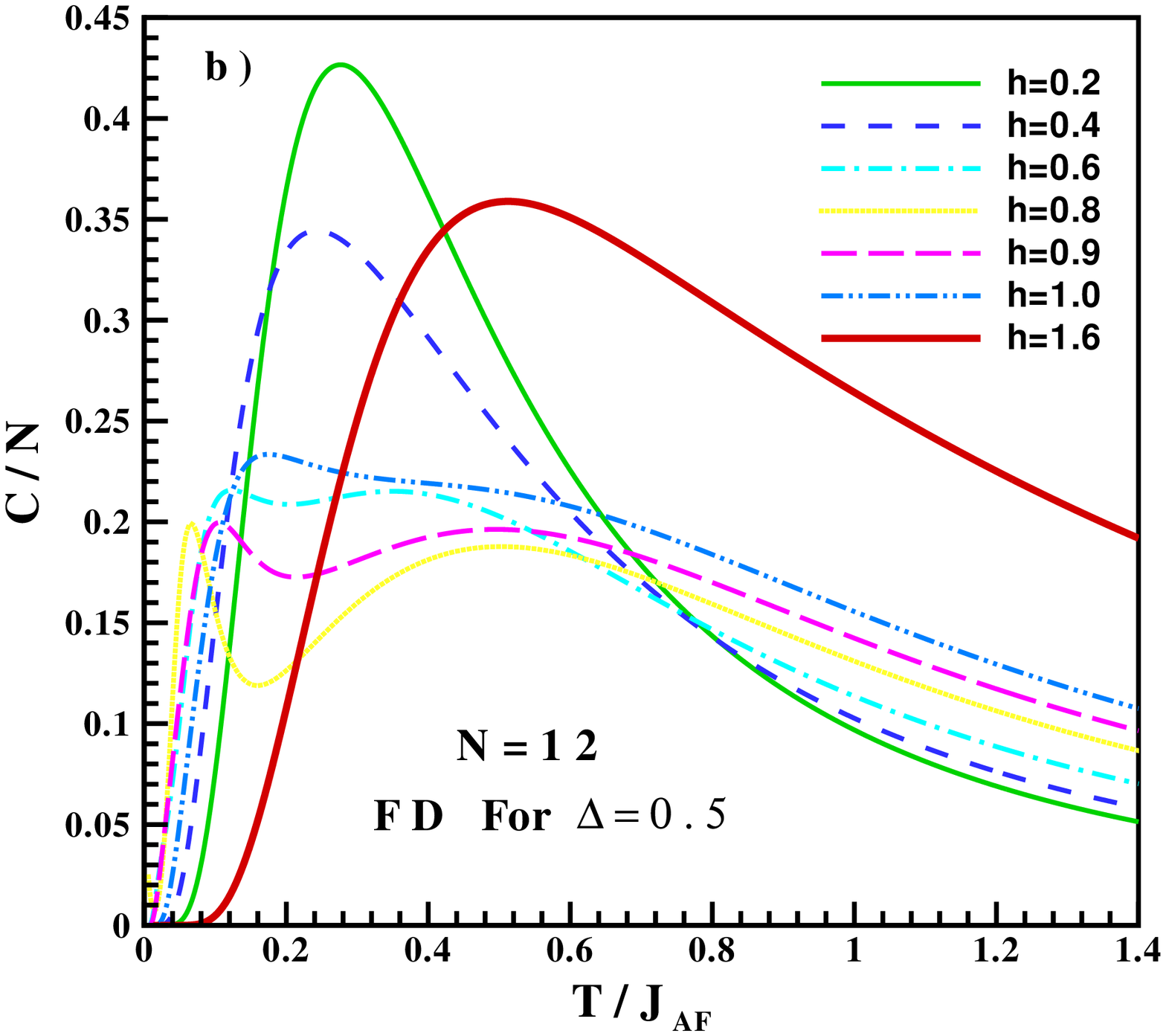}}
\caption{ Full diagonalization results of the specific heat of an
alternating AF-F spin-1/2 chain versus $T$ for different values of
the magnetic field $h$ with $J_{AF}=1$ and $J_{F}=0.5$. (a)
$\Delta=1$ (b) $\Delta=0.5$.} \label{fig6}
\end{figure}
\begin{figure}[t]
\centerline{\includegraphics[width=8cm,angle=0]{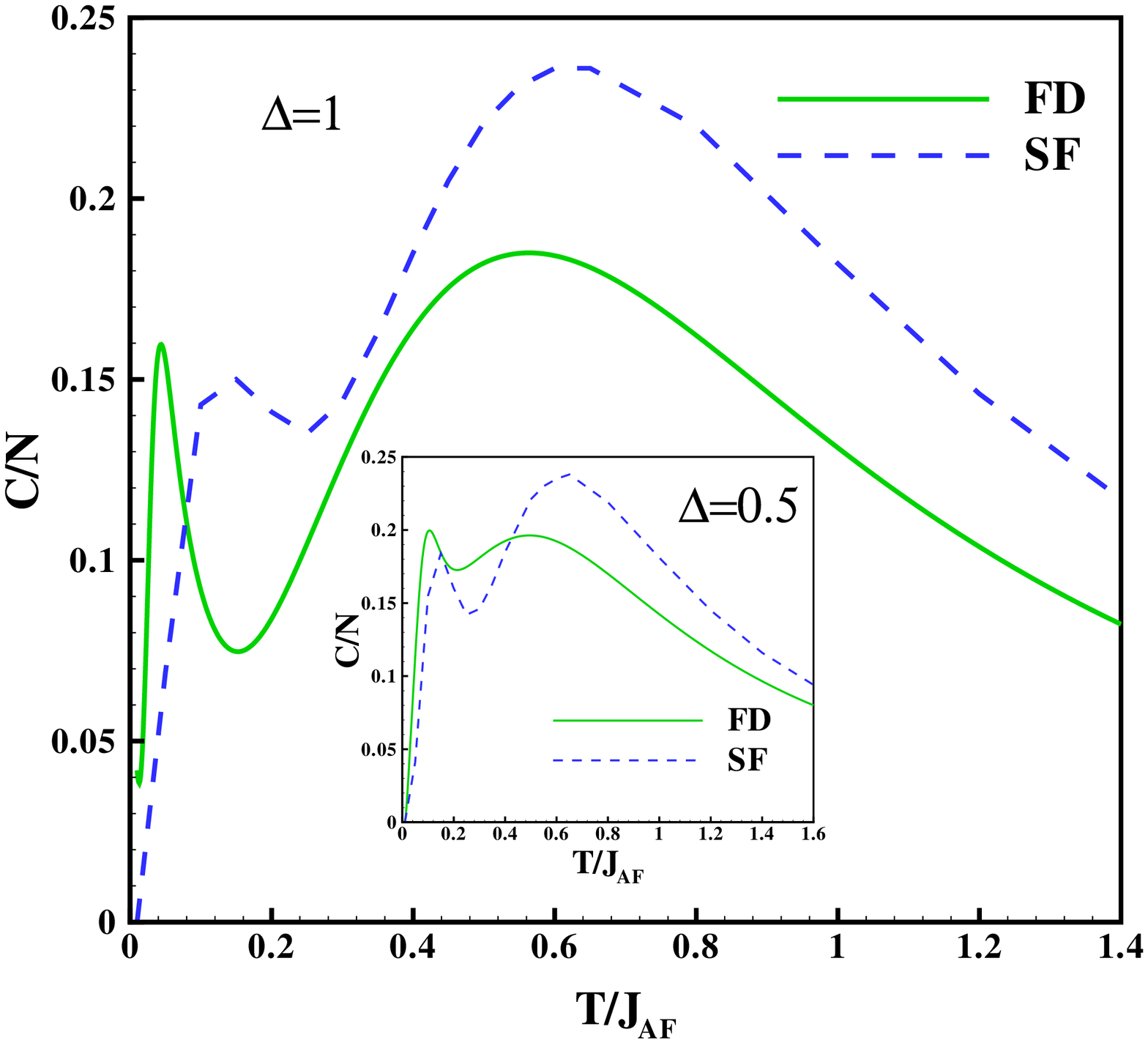}}
\caption{ Specific heat of the AF-F Heisenberg spin-1/2 chain vs. $T$ for
both values of $\Delta$ and $h=0.9$ using the spinless fermion (SF) and full diagonalization (FD) approaches.} \label{fig7}
\end{figure}
The discovery of gapless or gapped excitations have led to the
investigation of the thermodynamic properties of the model. One of
the most important thermodynamic functions is the specific heat.
In this respect, we study the temperature-dependence of the
specific heat of the model in different quantum regimes. We have
calculated the specific heat of both isotropic and anisotropic
cases, using two methods, the analytical SF and the numerical full
diagonalization methods. In Fig.~\ref{fig5}a(b) we have plotted
the results of the SF method in the form of the specific heat
$C/N$ versus $T$ with $\Delta=1$ and $\Delta=0.5$ for different
values of $h$. As it is clearly seen from this figure, in the
Haldane and paramagnetic regions ($h<h_{c_1}$ and $h>h_{c_2}$,
respectively), there is only one peak on the specific heat curve.
However, in the intermediate region ($h_{c_{1}}<h<h_{c_{2}}$) a
double peak is observed. At $h=0$ there is a Schottky like peak in
the specific heat which is expected from the Haldane phase. By
increasing $h$, the peak becomes wider and goes to lower
temperatures. Increasing $h$ further, causes a shoulder to appear
on the right hand side of the curve. For $h>h_{c_1}$, increasing
$h$, an additional peak appears which is the signal for the paramagnetic phase. Because of the two F and AF
interactions ($J_{F}$ and $J_{AF}$), there are two kinds of
quasi-particles in the system. These quasi-particles have two
different dynamics.
These dynamics bring about two energy scales in the system. The
energy scales affect the behavior of the response functions. It is
obviously seen from the plots of the specific heat $C/N$ that the
energy scales create a double peak in the specific heat at two
different temperatures. As a general statement, two energy scales
can produce a double peak structure in the specific heat.

We have also plotted in Figs. \ref{fig6} a(b) the specific heat of
the model by using the full diagonalization method. We have
computed all the eigenvalues of the energies for different values
of the transverse magnetic field $h$ and anisotropy parameter
$\Delta$. Therefore, using these eigen-energies, we have computed
the specific heat as a function of the temperature $T$. The
specific heat curves have been plotted for $J_{AF}=1$ and
$J_F=0.5$ and the anisotropy parameter $a)\Delta=1$,
$b)\Delta=0.5$. As it is clearly seen from this figures, in the
Haldane and paramagnetic regions ($h<h_{c_1}$ and $h>h_{c_2}$,
respectively), there is only one peak in the specific heat curve.
To see the qualitative agreement between spinless fermion results and Full diagonalization ones, we have plotted the specific heat versus $T$ for an intermediate value of $h$ (for example $h=0.9$). As it is seen from Fig.\ref{fig7} the mean field results are in agreement with the others, qualitatively.
It is interesting that the numerical results confirm the existence
of the double peak in the intermediate region
($h_{c_{1}}<h<h_{c_{2}}$).


\section{conclusion}
To summarize, we have studied the zero and
finite-temperature behavior of the anisotropic alternating AF-F
Heisenberg spin-1/2 chains in a transverse magnetic field. The
numerical exact diagonalization method and analytical spinless
fermion approach are applied to analyze the model. The first
notable point is introducing a new mean field order parameter
which can distinguish between a gapless LL phase and the gapped
phases. This order parameter in the spin language of Hilbert space
is the F-dimer order parameter. In the isotropic case, the F-dimer
order parameter has a considerable value in the LL region. We have
shown that there is a big discrepancy between the values of the
F-dimers of the isotropic and anisotropic cases, \emph{i.e.}, in
the LL gapless phase the behavior of the F-dimer order parameter
($ReP_F$) is different from that of the gapped phases. Therefore,
we have concluded that the F-dimer order parameter can distinguish
the gapless LL phase from the other gapped phases.

The second notable point is found from the specific heat. We have
obtained a double peak structure in the specific heat curves vs
temperature.

There are some questions behind the defined order parameter which are still controversial and we can investigate them in future. The most remarkable questions referred to the topological order and spontaneous symmetry breaking. \emph{I}.e.
in the LL phase, where the real part of the F-dimer is not zero, how can describe topological order and what kind of spontaneous symmetry breaking is occurred.


\section{Acknowledgments}
It is our pleasure to thank G. I. Japaridze for his valuable
comments and fruitful discussions. J. A. also would like to thank A. Langari and F. Shahbazi for their useful suggestions and comments.  We are also grateful to B. Farnudi for reading carefully the manuscript and appreciate his useful comments. J. A. was supported by the grant of Shahrood University of Technology.

\end{document}